# Observation of Chiral Surface State in Superconducting NbGe$_2$

Mengyu Yao,[1,*] Martin Gutierrez-Amigo,[2,3,4,†] Subhajit Roychowdhury,[1,†] Ion Errea,[5,6]
Alexander Fedorov,[7] Vladimir N. Strocov,[8] Maia G. Vergniory,[1,4,‡] and Claudia Felser[1,§]

[1]*Max Planck Institute for Chemical Physics of Solids, Dresden, Germany*
[2]*Departamento de Física, Facultad de Ciencia y Tecnología,*
*Universidad del País Vasco (UPV/EHU), Apartado 644, 48080 Bilbao, Spain*
[3]*Centro de Física de Materiales (CSIC-UPV/EHU),*
*Manuel de Lardizabal pasealekua 5, 20018 Donostia/San Sebastián, Spain*
[4]*Donostia International Physics Center, 20018 Donostia - San Sebastian, Spain*
[5]*Departamento de Aplicada, Facultad de Ciencia y Tecnología,*
*Universidad del País Vasco (UPV/EHU), Apartado 644, 48080 Bilbao, Spain*
[6]*Centro de Física de Materiales (CSIC-UPV/EHU),*
*Manuel de Lardizabal pasealekua 5, 20018 Donostia/San Sebastián, Spain*
[7]*Leibniz-Institut für Festkörper- und Werkstoffforschung Dresden e. V. Helmholtzstraße 20 01069 Dresden, Germany*
[8]*Photon Science Division, Paul Scherrer Institute, 5232 Villigen PSI, Switzerland*

(2024-03-16)

The interplay between topology and superconductivity in quantum materials harbors rich physics ripe for discovery. In this study, we investigate the topological properties and superconductivity of the nonsymmorphic chiral superconductor NbGe$_2$ using high-resolution angle-resolved photoemission spectroscopy (ARPES), transport measurements, and *ab initio* calculations. The ARPES data revealed exotic chiral surface states on the (100) surface originating from the inherent chiral crystal structure. Supporting calculations indicate that NbGe$_2$ likely hosts elusive Weyl fermions in its bulk electronic structure. Furthermore, we uncovered the signatures of van Hove singularities that can enhance many-body interactions. Additionally, transport measurements demonstrated that NbGe$_2$ exhibits superconductivity below 2K. Overall, our comprehensive results provide the first concrete evidence that NbGe$_2$ is a promising platform for investigating the interplay between non-trivial band topology, possible Weyl fermions, van Hove singularities, and superconductivity in chiral quantum materials.

The fascinating and intricate relationship between unique band topologies, the specific symmetries of crystal structures, and the phenomenon of superconductivity remains a highly active and intriguing area of research within the broad field of condensed matter physics and materials science [1]. The study of topological materials, characterized by protected metallic states on surfaces, the presence of unconventional fermionic particles, and a host of other quantum mechanical wonders, has captured the attention of the scientific community in recent years [2]. The introduction of superconductivity into systems that already exhibit topological novelty opens up a realm of potential for the discovery of Majorana particles, which are theorized to exist under such conditions. These particles, in turn, are fundamental to the ambitious pursuit of architectures for quantum computing that exploit topological principles for enhanced stability and functionality [3–8].

Against this backdrop, crystals that are chiral and lack symmetries such as inversion and reflection, classified under the category of nonsymmorphic chiral crystals, have captured the interest of researchers as a novel category of topological materials [9–13]. A particularly exciting theoretical proposition is that the bands of such crystals, influenced by spin-orbit coupling, are capable of hosting extraordinary fermionic excitations known as Kramers-Weyl fermions. These particles are predicted to occur at points of high symmetry within the crystal and are characterized by their topologically charged Weyl points, which are protected by the symmetry of the crystal [14]. The appeal of chiral crystals extends even further when these systems either naturally exhibit superconducting properties or are externally manipulated to become superconductors. In such cases, the superconducting state is expected to be topological in nature, with exotic features such as chiral Majorana modes located at the edges of the material [15].

Despite the tantalizing theoretical frameworks and predictions, the experimental identification and study of chiral crystals that are superconducting has been somewhat elusive. However, the landscape began to change with the recent spotlight on NbGe$_2$, a hexagonal transition metal dichalcogenide, as a potential chiral topological supercon-ductor. Although it was previously established that NbGe₂ becomes superconducting at temperatures below approximately 2K, the topological properties of this material and the detailed microscopic understanding of its superconducting phase had not been thoroughly investigated. Importantly, new theoretical insights suggest that NbGe₂ is a likely host for Kramers-Weyl fermions, attributed to its unique nonsymmorphic chiral crystal lattice [4–8]. In parallel, there have been reports indicating the presence of strong interactions between electrons and phonons in NbGe₂ [5]. These developments underscore the rich potential of NbGe₂ as a platform for exploring the interplay of chirality, topology, and superconductivity in a material system.

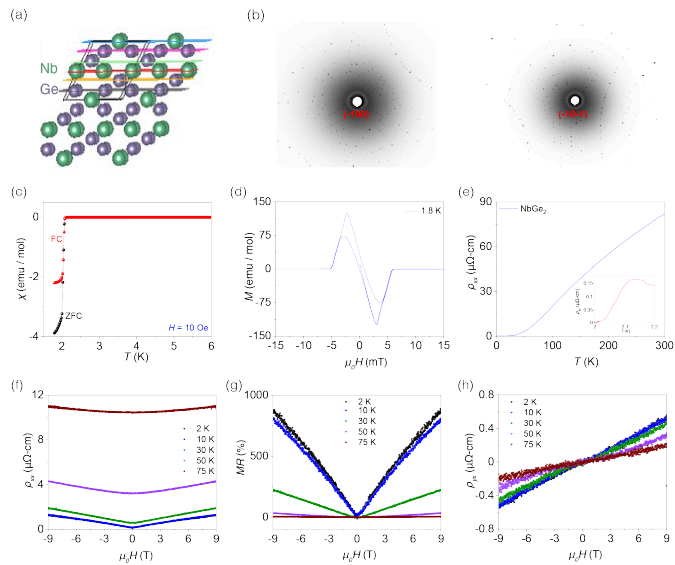

Figure 1: Schematics for NbGe₂ and sample characterization. (a) Crystal structure of NbGe₂. Possible terminations are indicated by the colored layers. (b) Laue diffraction patterns. (c) DC magnetic susceptibility ($\chi$) data as a function of temperature. (d) Magnetization as a function of magnetic field. (e) The resistivity ($\rho_{xx}$) as a function of temperature. Inset: Enlarged superconducting transition plot. Field dependent (f) resistivity ($\rho_{xx}$) (g) magnetoresistance, (h) Hall resistivity ($\rho_{yx}$) at different temperatures. Measurement configuration: I$_\perp$ [101], H$_\parallel$ [101] of NbGe₂ single crystal.

In this study, we employed a combination of experimental techniques, including angle-resolved photoemission spectroscopy (ARPES), transport measurements and density functional theory (DFT) calculations, to investigate the electronic properties of NbGe₂. Interestingly, while our ARPES and DFT studies reveal chiral-shaped surface states in NbGe₂, the transport measurements demonstrate no special features. This contrast presents an intriguing puzzle and suggests that the superconducting and topological properties of NbGe₂ may be more subtle than initially thought. The aim of this paper is to present our findings and discuss their implications for our understanding of superconductivity and chiral crystal structures.

Figure 1(a) illustrates the crystalline structure of bulk NbGe₂. NbGe₂ crystallizes in a non-centrosymmetric hexagonal structure, aligning itself with the $P6_222$ space group. Potential terminations of the (100) planes are illustrated in Figure 1(a), each distinguished by a uniquely colored layers. Notably, the termination of the measured samples exhibited temporal variability owing to the intrinsic characteristics of the sputtering/annealing cleaning procedure. Subsequent to each cleaning process, the termination invariably constitutes a conglomerate of all conceivable possibilities.

High quality single crystals of NbGe₂ were grown from a germanium (Ge) melt. The as-grown samples were characterized using Laue diffraction (Figure 1(b)). High-quality samples pave an important step for transport measurements, which reveal superconductivity (SC), and for electronic structure investigations using ARPES. We confirmed SC in NbGe₂ by measuring the dc magnetic susceptibility ($\chi$). Figure 1(c) illustrates the $T$-dependence of the magnetization. The data indicates that the diamagnetic transition temperature was 2.07K. A significant amount of vortex pinning is evident when the zero-field-cooled (ZFC) and the field-cooled (FC) data are compared below the transition temperature.

Figure 1(d) illustrates the isothermal magnetization loop at 2K. The hysteresis curve indicates typical type-II SC behavior. As presented in Figure 1(e), the longitudinal resistivity $\rho_{xx}$ decreases linearly with a decrease in temperature, which indicates a metallic nature of NbGe₂. Typically, the $\rho_{xx}$ value is approximately $0.15\mu\Omega \cdot$ cm at 2.2K and increases to $82\mu\Omega \cdot$ cm at 300K that is, the residual resistivity ratio RRR $= (\rho_{xx}(300K))/(\rho_{xx}(2.2K))$ value for NbGe₂ is 540, illustrating the high-quality single crystal. The SC transition temperature was consistent with the susceptibility measurements.

The field dependence of the longitudinal ($\rho_{xx}$) and Hall ($\rho_{yx}$) resistivities of the NbGe₂ crystal ($T = 2 - 75K$) is shown in Figure 1(f-h). Figure 1(g) illustrates the transverse magnetoresistance [MR $= (\rho_{xx}(\mu_0 H) - \rho_{xx}(0))/\rho_{xx}(0)$]. The MR attains its maximum value of approximately 900% at 2K and a magnetic field of 9T. However, it decays rapidly with increasing temperature. At low temperatures, the MR is quasilinear. At 100K, the MR becomes quadratic, reaching only 6% at 9T. This quasilinear behavior can be attributed to the presence of open Fermi surfaces as discussed previously by Ni et al. [8]. Figure 1(h) illustrates the field dependent Hall resistivity ($\rho_{yx}$) at different temperatures as a function of the magnetic field. The nonlinear Hall resistivity is observed due to the presence of both hole and electron pockets near the Fermi level. The carrier (hole) concentration of the single crystal is approximately $1.04 \times 10^{22}$cm$^{-3}$ at 2K.

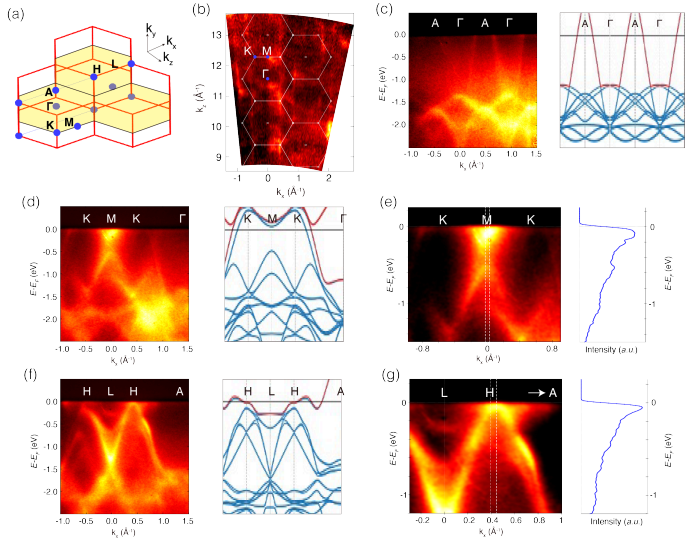

Figure 2: (a) Bulk Brillouin zone (BZ) of NbGe$_2$ with high-symmetry points labeled. (b) The out-of-plane Fermi surface (FS) mapping, taken with the photon energy in the range of 280-700 eV. The BZ boundary is marked by white lines. (c,d,f) Photoemission intensity plot and corresponding calculated band structure along high-symmetry directions. The high symmetry points are labeled on the figures. (e) Zoom-in view of (d). The density-of-state (DOS) around $M$ point is plotted in the right panel. (g) Shows the same information as (e), but around $H$ point.

In gain a comprehensive understanding of the electronic structure, we systematically performed bulk-sensitive ARPES experiments on NbGe$_2$ (100) with soft X-rays and surface-sensitive ones with UV light. Figure 2 shows the photon energy dependent measurements, which were used to construct Fermi surface mappings in the $k_x - k_z$ plane (Figure 2(b)). This mapping reveals a three-dimensional band dispersion, reflecting the inherent nature of the bulk bands in NbGe$_2$. Furthermore, we determined the photon energies using energy-dependent measurements and obtained several ARPES intensity plots along the corresponding high-symmetry directions (Figure 2(c,d and f)). These plots provide a detailed picture of the electronic band structures along these directions, which are particularly interesting because of their potential to host unique electronic states.

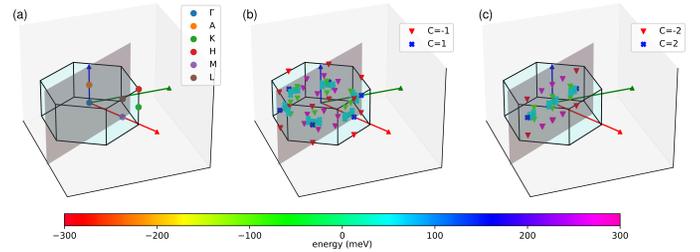

Figure 3: (a) Depiction of the Brillouin zone featuring its corresponding high-symmetry points. (b) Illustration of 84 Weyl points within the energy range of −300 to 300 meV, each exhibiting a Chern number with $|C| = 1$. (c) Projection of all Weyl points onto a plane parallel to the surface, effectively doubling the chiral charge of each Weyl point. The gray plane shown in all insets represents the (100) Miller plane, aligned parallel to the studied surface.

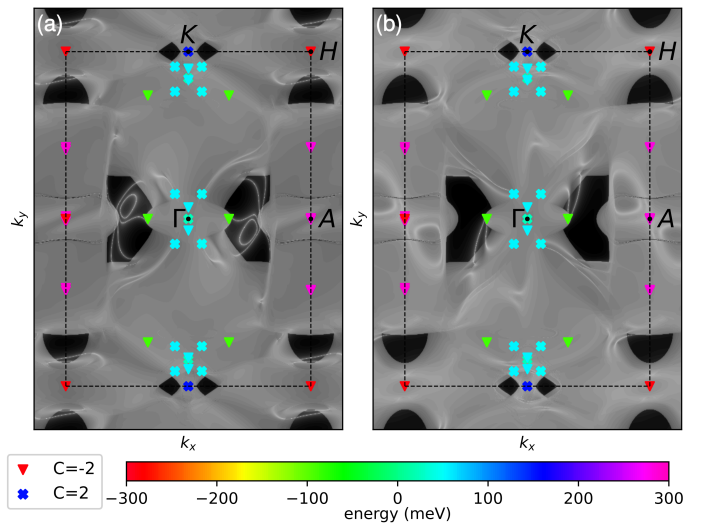

Figure 4: Surface density of states is illustrated for two distinct terminations (A and B) at an energy of 50 meV. Fermi arcs are discernible, appearing in pairs from the Weyl points located in proximity to the 50 meV energy range. Although the bulk Weyl points possess chiralities with $|C| = 1$, when projected onto a surface, they appear in pairs, resulting in an effective Chern number of $|C| = 2$.

To improve our understanding of the results, we compared the ARPES intensity plots with corresponding DFT calculations (Figure 2(c,d and f)). This comparison allowed us to confirm the accuracy of our measurements and provided theoretical insights into the observed electronic structure. We note that a saddle-point-like band dispersion is observed in both the ARPES data and DFT calculations close to $E_F$ around the $M$ and $H$ points, which indicates the presence of two vHSs. As illustrates in Figure 2(e,g), the existence of a vHS increases the local density-of-state (DOS). Interactions are enhanced [16–21], when the vHS is near $E_F$. In contrast, while previous theoretical research predicted the presence of Kramers-Weyl fermions [4], our current experimental resolution is insufficient to resolve the Weyl node predicted to be located at the $L$ point. However, our DFT calculations, however, support the theoretical prediction of multiple Weyl and Kramers-Weyl fermions across the Brillouin zone (Figure 3). A total of 84 Weyl points were identified just within the −300 to 300 meV energy range, each possessing chiralities of either 1 or −1. Even though the Chern number prediction of $|C| = 1$ suggests the appearance of a single Fermi arc, projecting all the identified Weyl points from the Brillouin zone onto the (100) Miller plane (parallel to the surface) reveals that they manifest in pairs (Figure 3(c)). This effectively doubles the chiral charge of all the Weyl and Kramers-Weyl fermions, leading to the emergence of Fermi arcs in pairs (Figure 4). Overall, the theoretical analysis underscores the need for future experiments with higher resolution to resolve the predicted Fermi arcs more accurately.

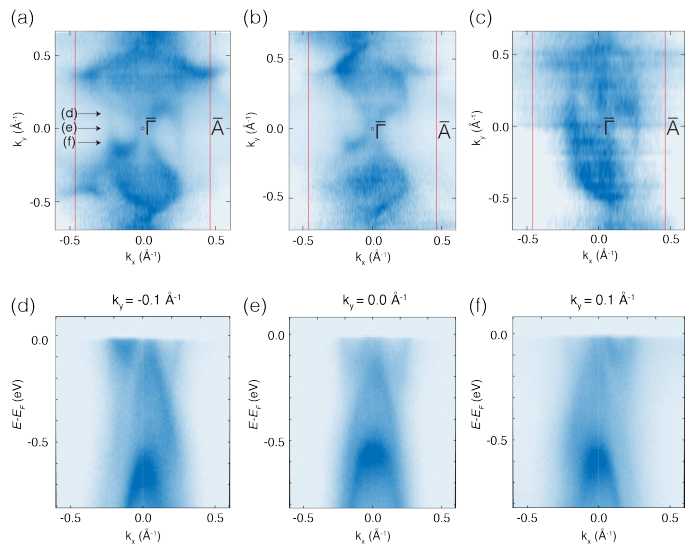

Figure 5: (a-c) The in-plane Fermi-surface (FS) mappings acquired with photon energies of 40, 50, and 120 eV, respectively. Brillouin zones (BZ) are indicated with red lines. (d-f) Photoemission intensity plots taken from (a) at $k_y$ = −0.1, 0 and 0.1 Å$^{-1}$, respectively.

Our UV-ARPES measurements reveal fascinating chiral fermion surface states in NbGe$_2$. Fermi surface mappings acquired at various photon energies (40 eV, 50 eV and 120 eV) all exhibited an intriguing chiral shape, despite differences in the photoemission intensity, as shown in Figure 5(a-c). Crucially, all the surface bands displayed an inversion symmetry with respect to the BZ center $(k_x, k_y) = (0, 0)$, which is a signature of their chiral nature. To analyze the band dispersions in detail, we extracted photoemission intensity plots along high symmetry directions from the 40 eV Fermi surface map (Figure 5(d-f)). At $k_y = -0.1 \text{Å}^{-1}$, the bands show an asymmetric dispersion about $k_x = 0$, which becomes mirrored at $k_y = 0.1 \text{Å}^{-1}$. This inversion symmetry between the two cuts directly reflects the two-fold rotation symmetry of the NbGe$_2$ crystal lattice on the (100) surface. Alongside the corroborating DFT simulations, our results provide compelling evidence that the observed exotic chiral fermion states originate from the intrinsic chiral crystal structure of NbGe$_2$. The discovered chiral surface states may engender unusual transport, superconducting and topological behaviors, motivating further investigations of this fascinating quantum material.

Our comprehensive study provides compelling evidence that NbGe$_2$ is an experimental platform that exhibits both chiral topological features and superconductivity. The combination of ARPES measurements directly observing exotic chiral surface states arising from the inherent lack of inversion and mirror symmetries, theoretical evidence for Weyl fermions in the bulk band structure, and transport measurements confirming superconductivity below 2K establishes NbGe$_2$ as a material that simultaneously hosts chirality, topology, and superconductivity.

The presence of chiral surface states in the NbGe$_2$ crystal structure is a remarkable discovery. These surface states exhibit a distinctive chiral shape and inversion symmetry, suggesting that they could lead to unusual transport phenomena. Meanwhile, the predicted bulk Weyl fermions, although not directly resolved in our current experiments, further enrich the topological character of the electronic structure of NbGe$_2$.

While our results do not directly establish a link between the chiral surface states and the superconducting phase, the coexistence of these topological features and superconductivity in NbGe$_2$ is intriguing. The Van Hove singularities observed close to the Fermi level may enhance electron correlations, potentially influencing the nature of the superconducting state. Further experimental and theoretical investigations are warranted to elucidate any interplay between topology and superconductivity in this material.

Overall, our work establishes NbGe$_2$ as a rare experimental realization that simultaneously exhibits chiral topological electronic features and superconductivity. Exploration of the phenomena arising from this unique combination could provide new insights into the rich physics of chiral quantum materials. Our results underscore the potential of nonsymmorphic chiral crystals as a fertile ground for discovering new quantum states and advancing the fundamental understanding of topology and superconductivity.

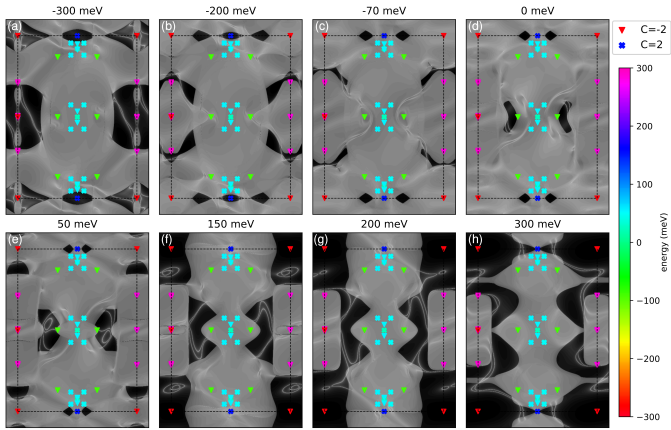

Figure 6: Surface density of states for termination A at different energies.

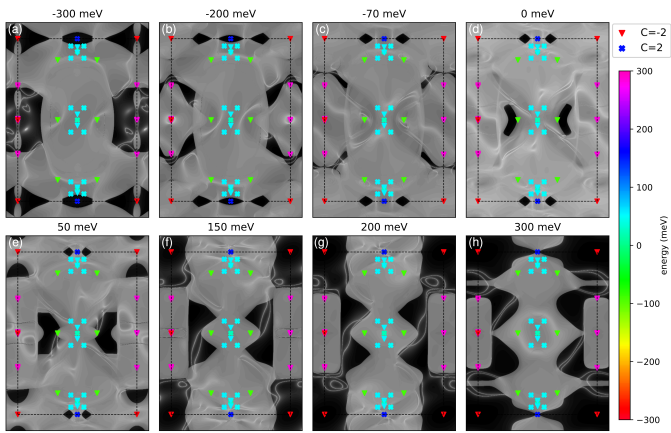

Figure 7: Surface density of states for termination B at different energies.

## Methods

- **ARPES Experiments**

For the ARPES experiments, the crystal surfaces were cleaned in-situ with multiple sputtering (Ar$^+$, 3keV, 30min, $4 \times 6$ mbar) and annealing ($> 600K$, $\gtrsim 30$ min). Soft X-ray (SX-ARPES) experiments were performed at the SX-ARPES endstation [22] of the ADRESS beamline [23] at the Swiss Light Source, Switzerland, using a SPECS analyzer with an angular resolution of 0.07°. The data were collected using photon energies in the soft X-ray region between 300 eV and 800 eV, with overall energy resolutions on the order of 50-80 meV [24]. Ultraviolet ARPES experiments were performed at the one-square endstate of the UE112-PGM beamline at the BESSY-II Synchrotron using a Scienta Omicron R8000 analyzer. All ARPES experiments were conducted with the sample maintained at approximately 20K and under a pressure better than $2 \times 10^{-10}$ mbar.

- **DFT Calculations**

First-principles density functional theory (DFT) calculations were conducted using the Quantum Espresso package [25,26]. We used the generalized gradient approximation with the Perdew-Burke-Ernzerhof parameterization [27] together with the projector-augmented wave pseudopotentials generated by Dal Corso [28]. In all cases, we used an energy cutoff of 60 Ry with a Methfessel-Paxton smearing [29] of 0.02 Ry. Structural relaxation was performed using a $12 \times 12 \times 10$ grid, without accounting for spin-orbit coupling (SOC) and was stopped when the pressures was below 0.01 kBar. Subsequently, SOC was included to compute the electronic band structures.

- **Energy contour plots**

For the energy contour plots, we used the Wannierization procedure implemented in Wannier90 [30], and WannierTools [31]. To generate the energy contour plots, we employed the Wannierization procedure implemented in Wannier90 [30], along with WannierTools [31]. First, we obtained a tight-binding model with a Wannierization considering $p$ and $d$ orbitals in the germananium and niobium sites. Subsequently, we calculated the surface spectrum using the iterative Green's function as implemented in WannierTools for a slab of 10 unit cells. Using the same tight-binding model, we employed WannierTools to detect and characterize the Weyl points along with their respective chiralities.

## Acknowledgement

This study was financially supported by an Advanced Grant from the European Research Council (No. 742068) 'TOPMAT', the European Union's Horizon 2020 research and innovation programme (No. 824123) 'SKYTOP', the European Union's Horizon 2020 research and innovation programme (No. 766566) 'ASPIN', the Deutsche Forschungsgemeinschaft (Project-ID 258499086) 'SFB 1143', the Deutsche Forschungsgemeinschaft (Project-ID FE 633/30-1) 'SPP Skyrmions', the DFG through the Würzburg-Dresden Cluster of Excellence on Complexity and Topology in Quantum Matter ct.qmat (EXC 2147, Project-ID 39085490). M.G.V. and M.G.A. thanks the Ministry for Digital Transformation and of Civil Service of the Spanish Government through the QUANTUM ENIA project call - Quantum Spain project, and by the European Union through the Recovery, Transformation and Resilience Plan - NextGenerationEU within the framework of the Digital Spain 2026 Agenda. M.G.V., M.G.A. and I.E. acknowledge the Spanish Ministerio de Ciencia e Innovacion (grants PID2019- 109905GB-C21, PID2022-142008NB-I00, and PID2022-142861NA-I00). I.E. acknowledges the Department of Education, Universities and Research of the Eusko Jaurlaritza and the University of the Basque Country UPV/EHU (Grant No. IT1527-22). M.G.V. and C.F. thanks support to the Deutsche Forschungsgemeinschaft (DFG, German Research Foundation) GA 3314/1-1 – FOR 5249 (QUAST) and partial support from European Research Council (ERC) grant agreement no. 101020833.

We acknowledge Prof. Dr. J. Fink for his insightful discussions. We acknowledge Dr. Maxim Krivenkov, Dr. Procopi Constantinou and Enrico Della Valle for their help with ARPES experiments.